\documentclass[12pt,a4paper]{article}

\usepackage{cite}
\usepackage{graphicx} 
\usepackage{caption} 
\usepackage{subcaption} 
\usepackage{float} 
\usepackage{setspace} 
\usepackage{pgf}

\usepackage{amsmath} 
\usepackage{amssymb} 
\numberwithin{equation}{section}

\usepackage[utf8]{inputenc}       
\usepackage[T1]{fontenc} 

\usepackage[english]{babel}           

\usepackage{tabularx} 

\usepackage{psfrag} 
\usepackage{sistyle} 

\usepackage{eurosym} 
\def\{\euro{}}

\usepackage{color} 

\usepackage{epstopdf}

\definecolor{linkcolor}{rgb}{0,0,0.6} 
\usepackage[ pdftex,colorlinks=true,
pdfstartview=FitV,
linkcolor= linkcolor,
citecolor= linkcolor,
urlcolor= linkcolor,
hyperindex=true,
hyperfigures=false]
{hyperref} 

\usepackage{fancyhdr} 

\usepackage{slashed} 

\usepackage{fancybox}

\usepackage{setspace}
\usepackage{newunicodechar}
\newunicodechar{ﬁ}{fi}
\newunicodechar{ﬀ}{ff}

\usepackage{tikz}
\usepackage{pgfplots}
\usepackage{xparse}

\usetikzlibrary{positioning,arrows,patterns}
\usetikzlibrary{decorations.markings}
\usetikzlibrary{calc}
\usetikzlibrary{shapes.arrows}
\usepackage{pifont} 
\tikzset{
  photon/.style={decorate, decoration={snake}, draw=black},
  fermion/.style={draw=black, postaction={decorate},decoration={markings,mark=at position .55 with {\arrow{>}}}},
  vertex/.style={draw,shape=circle,fill=black,minimum size=3pt,inner sep=0pt},
  scalar/.style={dashed,draw=black, postaction={decorate}},
  gluon/.style={decorate, draw=black,decoration={coil,amplitude=4pt, segment length=5pt}},
  plain/.style={draw=black},
}

\makeatletter
\@ifpackageloaded{babel}%
        {}
        {}
\makeatother

\usepackage{mathtools}

\title{Stage M2}

\author{Lilian Chabrol}

\begin{document}
\begin{titlepage}
		\begin{center}
\rightline{\small }

\begin{flushright} 
IPhT-T19/002
\end{flushright}

\vskip 2cm
\end{center}

	\begin{center}
			\LARGE Geometry of $\mathbb{R}^{+}\times E_{3(3)}$ Exceptional Field Theory and F-theory
	\end{center}
	\begin{center}
	{ Lilian Chabrol }
\vskip 0.1cm
{\small\it   Institut de Physique Th\'eorique, 
Universit\'e Paris Saclay, CEA, CNRS\\
Orme des Merisiers \\
91191 Gif-sur-Yvette Cedex, France} \\
\vskip 0.4cm
lilian.chabrol@ipht.fr
\vskip 0.8cm
	\end{center}
	\begin{center}
		\textbf{Abstract}
	\end{center}
	\vspace{0.2cm}We consider a non trivial solution to the section condition in the context of $\mathbb{R}^{+}\times E_{3(3)}$ exceptional field theory and show that allowing fields to depend on the additional stringy coordinates of the extended internal space permits to describe the monodromies of (p,q) 7-branes in the context of F-theory. General expressions of non trivial fluxes with associated linear and quadratic constraints are obtained via a comparison to the embedding tensor of eight dimensional gauged maximal supergravity with gauged trombone symmetry. We write an explicit generalised Christoffel symbol for $E_{3(3)}$ EFT and show that the equations of motion of F-theory, namely the vanishing of a 4 dimensional Ricci tensor with two of its dimensions fibered, can be obtained from a generalised Ricci tensor and an appropriate type IIB ansatz for the metric.
	
	\noindent

\vfill

\today

	\end{titlepage}

\tableofcontents
\newpage
\section{Introduction}
	
String theory shows a particularly rich structure of discrete duality symmetries after toroidal compactification given by the discrete split real forms of exceptional groups $E_{d(d)}(\mathbb{Z})$, where d is the dimension of the compactification space. In its low energy limit, those discrete groups become continous and 11 dimensional supergravity has a global $E_{d(d)}(\mathbb{R})\equiv E_{d(d)}$ symmetry group, whereas for type II supergravity theories the symmetry groups are $E_{d+1(d+1)}$. In the case of T-duality this led to the construction of Double Field Theory (DFT) \cite{hull_double_2009,hull_gauge_2009,hohm_background_2010,hohm_generalized_2010} as well as Generalised Geometry \cite{hitchin_generalized_2003, gualtieri_generalized_2004}, which make manifest an O(d,d) symmetry. DFT was constructed using a doubled space with associated additional "winding coordinates" \cite{hull_geometry_2005,hull_doubled_2007,dabholkar_generalised_2006,hull_gauge_2008,hull_flux_2006} . They are later removed by a \textit{section condition} to recover a physical theory. Generalised geometry on the other hand extends the tangent space $T$ to the combination $T\oplus T^{*}$, thus describing both vectors and 1-forms in a unique fiber. Both theories are manifestly O(d,d)-covariant, and combine diffeomorphisms as well as B-field gauge transformations in a single object: double vectors in DFT and sections of the generalised fiber in generalised geometry. Extensions of those theories were constructed to consider the full U-duality and the expected $E_{d(d)}(\mathbb{R})$ symmetry one gets from string theory compactifications in the low energy limit: Exceptional Field Theory (EFT) \cite{berman_duality_2012,berman_local_2012,berman_generalized_2011,thompson_duality_2011} and Exceptional Generalised Geometry (EGG)\cite{hull_generalised_2007,pacheco_m-theory_2008}. The group of symmetry is larger when one considers the S-duality in addition to T-duality. Thus, the space is no longer doubled for exceptional field theories but is rather decomposed into an external space and an extended internal one. The geometric structure of this internal space is then constructed to be manifestly $E_{d(d)}$ covariant in order to render manifest the symmetries between the NS-NS and RR fields after compactification. Generalised vectors on this extended internal space and sections of the generalised fiber in the case of EGG then describe usual diffeomorphisms combined with NS-NS and RR gauge transformations \cite{coimbra_$e_dd$_2011,coimbra_supergravity_2012,hohm_unification_2011,hohm_double_2011, coimbra_supergravity_2011,berman_gauge_2013}. In this context, massless type II and eleven dimensional gauged supergravities were obtained in a unified framework in various dimensions\cite{hohm_exceptional_2014,hohm_exceptional_2014-1,hohm_exceptional_2014-2,hohm_tensor_2015,abzalov_exceptional_2015,musaev_exceptional_2016,berman_action_2016}. Massive type IIA was then obtained using a violation of the section condition in double field theory \cite{hohm_massive_2011} as well as deformation of the generalised Lie derivative structure in the context of EFT and EGG \cite{ciceri_exceptional_2016,cassani_exceptional_2016}. All of this leads to wonder if EFT or EGG permits to describe aspects of generalisation of massless type IIB string theory, and in particular F-theory. Its link to $SL(2)\times \mathbb{R}^{+}$ EFT was first looked at in \cite{berman_action_2016}, and we ourself will focus on $SL(3)\times SL(2)\times \mathbb{R}^{+}$ EFT which, with an appropriate ansatz, contains both NS-NS and RR two-forms.

	F-theory can be thought of as a geometric formulation of type IIB string theory with D7 branes, and more generally (p,q) 7-branes \cite{vafa_evidence_1996,bergshoeff_q7-branes_2007}\footnote{For comprehensive reviews on F-theory see for example  \cite{blumenhagen_basics_2010,weigand_lectures_2010,weigand_tasi_2018}.}. The axion ($C_{0}$) and the dilaton ($\phi$) of type IIB string theory are combined into a complex field, the axio-dilaton ($\tau = C_{0}+ie^{-\phi}$), which becomes the complex parameter of a fibered 2-torus with constant volume fibered over a 10 dimensional space, thus leading one to consider a pseudo 12 dimensional theory. As the D7 brane is a magnetic charge for the axion, the axio-dilaton presents a singularity at the location of the brane, leading to a monodromy as one goes around it. Now, considering the type IIB effective action, one shows that the monodromies needs not only act on the axio-dilaton but also on the NS-NS and RR two forms $B_{2}$ and $C_{2}$ respectively. In the general case of (p,q) 7-branes the monodromies are entirely described by $SL(2, \mathbb{Z})$, and they act on the NS-NS and RR two-forms fields by mixing them. As one combines gauge transformations of those fields in a single object in EFT and EGG one can ask whether they can also describe monodromies and type IIB supergravity with D7 branes.\\
	
	Here we focus on the $E_{3(3)}= SL(3)\times SL(2)$ exceptional field theory arising for compactifications of type IIB to 8 dimensions \cite{hohm_tensor_2015}. In order to describe properly warped compactification of type IIB supergravity one has to consider that the duality group is extended  by a conformal factor to $\mathbb{R}^{+}\times E_{3(3)}$. In the context of supergravity the conformal symmetry is called trombone symmetry and can be thought of as a generalisation of the rescaling symmetry of the metric in Einstein's theory of gravity \cite{cremmer_spectrum-generating_1997,diffon_supergravities_2009}.
 In the first part of this paper we present a review of the basic results of $\mathbb{R}^{+}\times SL(3)\times SL(2)$ EFT and in particular the sufficient conditions needeed for a consistent theory. We then compute the fluxes of the theory, which we compare to the embedding tensor of the associated supergravity theory with gauged trombone symmetry. As the gauging of the trombone symmetry was only done for simple groups, we present a construction in the particular case where the original global group symmetry is $SL(3)\times SL(2)$. We then construct explicitely a generalised Christoffel symbol and remind the reader about the construction of a generalised Ricci tensor done in \cite{aldazabal_extended_2013}. We then focus on a non standard solution to the section condition leading to the description of the monodromies of (p,q) 7-branes in F-theory. This is done by considering that the fields of the final theory will have a dependency on 2 coordinates of the internal extended space, which are linear combination of both the usual coordinates and the stringy coordinates. This in particular ensures that product and inverse of fields have a similar dependency on the generalised space and are therefore also solutions to the section condition. The description of the monodromies leads to the breaking of both gauge transformations of $B_{2}$ and $C_{2}$ which seem to be entirely constrained. This is however not a particular issue as the monodromies of (p,q) 7-brane where only constructed when the only non zero field living on the brane was $C_{8}$, the dual field of the axion $C_{0}$. It is thus plausible that when one is describing the full backreaction of the brane with non trivial NS-NS and RR fields living on its world volume, the gauge symmetry of these fields normal to the brane would be broken. Finally, when one considers the standard solution to the section condition, we show that the generalised Ricci tensor gives the equations of motion of F-theory as a Ricci-flatness of a four dimensional space with two fibered directions.
\section{Structure of $SL(3)\times SL(2)$ Exceptional Field Theory}
	
		Compactifying M-theory on a d-dimensional torus, or Type II on a d-1 torus leads to an underlying U-duality symmetry given by the exceptional groups $E_{d(d)}(\mathbb{Z})$. In the low energy limit where we recover the eleven dimensional and massless type II supergravities, an underlying $E_{d(d)}(\mathbb{R})$ global symmetry appears. This symmetry can be made manifest in the context of exceptional field theory where the space is decomposed into an external space, and an internal extented space. The structure of the internal space will then differ from Riemannian geometry and have a manifest $E_{d(d)}$ covariance. Here we will consider d=3, corresponding to a 8-dimensional external space combined to a 6-dimensional internal extended space with a $E_{3(3)} = SL(3)\times SL(2)$ geometric structure. In fact, as mentioned in the introduction, one can extend this duality group by considering the trombone symmetry appearing in supergravity theories. The duality group becomes therefore $\mathbb{R}^{+}\times SL(3)\times SL(2)$. The extended internal space will be our main focus throughout this paper as the tensor hierarchy of $E_{3(3)}$ exceptional field theory is done in \cite{hohm_tensor_2015}. We now present the basics of $\mathbb{R}^{+}\times SL(3)\times SL(2)$ exceptional field theory which will be needed throughout this paper. \\
		
	We introduce a set of coordinates $X^{M}$, where M,N,P = 1,..,6 of the 6-dimensional internal space which lives in the vector representation \textbf{(3,2)} of $SL(3)\times SL(2)$. We can decompose the index of the fundamental representation M into $M = m \gamma$ where all Latin letters $m, n, p,... =1,2,3$ and all Greek letters $\gamma, \eta, \rho,...=1,2$ correspond respectively to the $SL(3)$ and $SL(2)$ part of $E_{3 (3)}$. We will note $\partial_{M} = \partial_{m \gamma}$ the derivative with respect to $X^{M} = X^{m \gamma}$.\\

	The usual action of Riemannian Lie derivatives does not preserve the group structure of the theory: this leads to the introdution of a generalised Lie derivative \cite{coimbra_$e_dd$_2011} which can be written for any exceptional geometry as \cite{berman_gauge_2013}
	\begin{equation}
	\label{diffZ}
		\mathcal{L}_{\Lambda} V^{M} = \mathbb{L}_{\Lambda}V^{M} + Z^{M N}{}_{P Q} \partial_{N}\Lambda^{P}V^{Q}+\left(\lambda(V)-\frac{1}{6})\right)\partial_{N}{\Lambda^{N}}V^{M}
	\end{equation}
	where $\mathbb{L}$ is the usual Riemannian Lie derivative and $\lambda(V)$ the conformal weight of the vector V. The tensor Z encodes the deviation from Riemannian geometry and is given in terms of the invariants of the duality group, which in our case is
	\begin{equation}
	Z^{M N}{}_{P Q} = Z^{m \gamma n \eta}{}_{p \rho q \sigma} = \epsilon^{m n z}\epsilon_{p q z} \epsilon^{\gamma \eta}\epsilon_{\rho \sigma}
	\end{equation}
	 where $\epsilon$s are totaly antisymmetric invariant tensors of $SL(3)$ and $SL(2)$. The invariant tensor verifies in particular $\mathcal{L}Z = 0$. Another expression for the generalised Lie derivatives which will be useful later to determine the fluxes of the extended space is
	\begin{align}
	\begin{split}
	\label{diffProj}
	\mathcal{L}_{\Lambda}V^{M} = \Lambda^{N}\partial_{N}V^{M} - 2(\mathbb{P}_{\mathbf{(8,1)}})^{M}{}_{N}{}^{P}{}_{Q} \partial_{P}\Lambda^{Q} V^{N} - 3(\mathbb{P}_{\mathbf{(1,3)}})^{M}{}_{N}{}^{P}{}_{Q} \partial_{P}\Lambda^{Q} V^{N}&\\
	+\lambda(V)\partial_{N}\Lambda^{N} V^{M}&
	\end{split}
	\end{align}
	where $\mathbf{(8,1)\oplus (1,3)}$ is the adjoint of $SL(3)\times SL(2)$ and the projections on each subspaces are given by
		\begin{align}
		\begin{split}
		\label{projectors}
			(\mathbb{P}_{\mathbf{(8,1)}})^{M}{}_{N}{}^{P}{}_{Q} = (\mathbb{P}_{\mathbf{(8,1)}})^{m \gamma}{}_{n \eta}{}^{p \rho}{}_{q \sigma} = \frac{1}{2}\delta^{\gamma}_{\eta}\delta^{\rho}_{\sigma}\left(\delta_{n}^{p}\delta_{q}^{m} - \frac{1}{3}\delta_{n}^{m}\delta_{q}^{p}\right) = \frac{1}{2}\delta^{\gamma}_{\eta}\delta^{\rho}_{\sigma}\left(\mathbb{P}_{\mathbf{8}}\right)^{m}{}_{n}{}^{p}{}_{q}\\
			(\mathbb{P}_{\mathbf{(1,3)}})^{M}{}_{N}{}^{P}{}_{Q} = (\mathbb{P}_{\mathbf{(1,3)}})^{m \gamma}{}_{n \eta}{}^{p \rho}{}_{q \sigma} = \frac{1}{3}\delta^{m}_{n}\delta^{p}_{q}\left(\delta_{\eta}^{\rho}\delta_{\sigma}^{\gamma} - \frac{1}{2}\delta_{\eta}^{\gamma}\delta_{\sigma}^{\rho}\right) = \frac{1}{3}\delta^{m}_{n}\delta^{p}_{q}\left(\mathbb{P}_{\mathbf{3}}\right)^{\gamma}{}_{\eta}{}^{\rho}{}_{\sigma}
			\end{split}
		\end{align}
	with $\mathbb{P}_{\mathbf{8}}$ and $\mathbb{P}_{\mathbf{3}}$ the projectors onto the $SL(3)$ and $SL(2)$ adjoint respectively. The expressions of the projectors onto the adjoint using the generators of $SL(3)$ and $SL(2)$ are detailed in Appendix \ref{Projector}. Finally, using \eqref{projectors} we can write the generalised Lie derivative in terms of $SL(3)$ and $SL(2)$ indices  
	\begin{equation}
		\mathcal{L}_{\Lambda}V^{m \gamma} = \Lambda^{n \eta}\partial_{n \eta}V^{m \gamma}-V^{m \eta}\partial_{n \eta}\Lambda^{n \gamma} - V^{n \gamma}\partial_{n \eta}\Lambda^{m \eta} + \left(\lambda(V)+\frac{5}{6}\right)\partial_{n \eta}\Lambda^{n \eta} V^{m \gamma}.
	\end{equation}

In order for the theory to be consistent, the algebra of the generalised Lie derivatives \eqref{diffZ} has to close, i.e. it should satisfy
		\begin{equation}
			\left[\mathcal{L}_{\Lambda_{1}}, \mathcal{L}_{\Lambda_{2}}\right] = \mathcal{L}_{\Lambda_{12}}
		\end{equation}
		where
	\begin{equation}
		\Lambda_{12} \equiv [\Lambda_{1},\Lambda_{2}]_{E} = \frac{\mathcal{L}_{\Lambda_{1}}\Lambda_{2} - \mathcal{L}_{\Lambda_{2}}\Lambda_{1}}{2}
	\end{equation}
	is the analogy of the Courant bracket introduced in generalised geometry but in the context of exceptional geometry \cite{pacheco_m-theory_2008,coimbra_$e_dd$_2011}. The closure of the algebra however is only achieved if one imposes a constraint on the different fields known as the section condition\footnote{More generally, there are four constraints which in the case of the split forms of the exceptional groups $E_{d(d)}$ (d=2..7) are equivalent to the section condition \cite{berman_gauge_2013}.} 
	\begin{align}
	\begin{split}
	\label{strongConstraint} Z^{NK}{}_{PQ} \partial_{N}\otimes\partial_{K} = \epsilon^{n k z}\epsilon_{p q z} \epsilon^{\eta \kappa}\epsilon_{\rho \delta}\partial_{n \eta}\otimes \partial_{k \kappa} = 0& \\
	 \Leftrightarrow \partial_{n \eta}\otimes \partial_{k \kappa} - \partial_{n \kappa}\otimes \partial_{k \eta} + \partial_{k \kappa}\otimes \partial_{n \eta} - \partial_{k \eta}\otimes \partial_{n \kappa} = 0&.
	 \end{split}
\end{align}
The fields of the theory therefore can no longer depend arbitrarily on the 6 dimensional internal space, but in our case rather a 2 or 3 dimensional subspace. This allows one to describe in particular 8+3=11 dimensional supergravity or 8+2=10 dimensional type II supergravity respectively.
We will consider the embedding of type IIB supergravity: we will focus on the solutions where the fields effectively depend on a two dimensional subspace of the six dimensional internal space. The usual way to do this is to consider that
\begin{equation}
	\label{trivialAssump}
	\partial_{1 \gamma}(A) =  \partial_{2 \gamma}(A) = 0
\end{equation} 
for any field A.
This leads to the breaking of $SL(3)$ into $SL(2)\times U(1)$. To make this breaking manifest we can split the index $M = m \gamma$ of the fundamental representation into $m \gamma = \left(\hat{m} \gamma, 3 \gamma\right)$ where $\hat{m}=1,2$.  
 \section{Fluxes}

 Compactifying string theory with fluxes leads, in the low energy limit, to gauged supergravity. They correspond to deformation of abelian supergravities where a subgroup $G_{0}$ of the global symmetry group G of the supergravity theory is promoted to a local symmetry. The embedding of the gauge group $G_{0}$ into the global symmetry group G can be described by an object called the embedding tensor, which corresponds exactly to the fluxes. Supersymmetry and gauge invariance of the embedding tensor then leads to a set of linear and quadratic constraint on the embedding tensor, which by extension should be verified by the fluxes of the corresponding low energy limit of string theory \cite{samtleben_lectures_2008, de_wit_lagrangians_2003}. 
 
 Here we derive the expression of the generalised fluxes for the $\mathbb{R}^{+}\times SL(3)\times SL(2)$ exceptional field theory. They will have to verify both linear and quadratic constraints so that the corresponding 8 dimensional gauged maximal supergravity we obtain in the low energy limit after compactification with fluxes is consistent. Considering the warp factor in the duality group will lead us to consider gauged supergravity with a gauged trombone symmetry. The gauging of the trombone symmetry for $SL(3)\times SL(2)$ exceptional field theory has never been done before due to the group product structure of this particular theory. We construct it here similarly to was done in \cite{diffon_supergravities_2009} where the trombone symmetry was gauged for simple groups.

\subsection{Embedding tensor structure of D=8 Gauged Maximal Supergravity with trombone symmetry}

	A way to describe the gauging of a subgroup of a global symmetry group G in supergravity theories is through the constant embedding tensor $\Theta_{M}{}^{\Gamma}$ \cite{samtleben_lectures_2008,de_wit_lagrangians_2003}, where M in our case corresponds to the fundamental representation $\mathbf{(3,2)}$ and $\Gamma$ an index of $\mathbf{Adj(G) = Adj\left(SL(3)\times SL(2)\right)}$. A consistent local gauging of the theory forces one to consider two constraints on this embedding tensor: a linear one and a quadratic one. Let us recall the results already known for the particular case of $E_{3(3)}$ exceptional field theory, without the scale factor of the general extended group. A priori the embedding tensor $\Theta_{M}{}^{\Gamma}$ of the theory lives in
	\begin{equation}
		\mathbf{(3,2)\times ((8,1) +(1,3)) = \left[(3,2) + (6,2) + (15,2)\right]+\left[(3,2) + (2,4)\right]]}
	\end{equation}
but due to the linear and quadratic constraints, the embedding tensor only has $\mathbf{(6,2)}$ and $\mathbf{(3,2)}$ components. Using this linear constraint we can write the generators of the gauge group of the theory using the embedding tensor and the generators of the adjoint of the gauge group $\lbrace t_{\Gamma} \rbrace$
\begin{equation}
	\label{gaugeGen}
	(X_{M})_{N}{}^{P} = \Theta_{M}{}^{\Gamma}(t_{\Gamma})_{N}{}^{P} = \Theta_{m \gamma, n}{}^{p}\delta_{\eta}^{\rho} + \Theta_{m \gamma, \eta}{}^{\rho}\delta_{n}^{p}
\end{equation}
with
\begin{align}
	\begin{split}
	\Theta_{m \gamma, \eta}{}^{\rho} = \xi_{m \eta} \delta_{\gamma}^{\rho} - \frac{1}{2}\delta_{\eta}^{\rho}\xi_{m \gamma} = \mathbb{P}_{\mathbf{(3_{SL(2)})}}{}^{\rho}{}_{\eta}{}^{\delta}{}_{\gamma} \xi_{m \delta}\\
	\Theta_{m \gamma, n}{}^{p} = f_{\gamma}{}^{(p b)} \epsilon_{b m n} + x(\xi_{n \gamma} \delta_{m}^{p} - \frac{1}{3} \xi_{m \gamma} \delta_{n}^{p}) = f_{\gamma}{}^{(p b)} \epsilon_{b m n} + x \mathbb{P}_{\mathbf{(8)}}{}^{p}{}_{n}{}^{r}{}_{m} \xi_{r\gamma}
	\end{split}
\end{align}
where the constant x is equal to $-\frac{3}{4}$, $\Theta_{m \gamma, \eta}{}^{\rho}\delta_{n}^{p} \in \mathbf{(3,2)}$ and $\Theta_{m \gamma, n}{}^{p}\delta_{\eta}^{\rho}\in \mathbf{(6,2)}$\footnote{$f_{\gamma}{}^{m n}$ and $\xi_{m \alpha}$ need to verify a set of quadratic constraints which can be found in \cite{de_roo_critical_2012}.}. To avoid confusion between the fundamental representation of $SL(3)$ and the adjoint of $SL(2)$ we denoted the later $\mathbf{(3_{SL(2)})}$.\\

This is not the more general story of supergravity gauging however, as one can gauge the trombone symmetry \cite{cremmer_spectrum-generating_1997,diffon_supergravities_2009}. In order to do that we have to consider a more general ansatz than the one used in \cite{diffon_supergravities_2009}, as the global symmetry group is not simple in our case but a product of simple groups. Considering the $\mathbb{R}^{+}$ factor in the duality group leads to an additional generator $(t_{0})_{N}{}^{P} = -\delta_{N}^{P}$ in equation \eqref{gaugeGen}, and a corresponding additional component of the embedding tensor $\Theta_{M}{}^{0} \equiv K_{M}$. This component lives in the \textbf{(3,2)} representation, and we expect it to appear in the same way as the other \textbf{(3,2)} parameter $\xi_{M}$. This leads to the following ansatz for the generators of the gauge group
\begin{align}
\label{gaugeGenTrom}
	X_{MN}{}^{P} =& \Theta_{m \gamma, n}{}^{p} \delta_{\eta}^{\rho} + \Theta_{m \gamma, \eta}{}^{\rho} \delta_{n}^{p} + \left(\zeta_{1}\mathbb{P}_{\mathbf{(8)}}{}^{p}{}_{n}{}^{k}{}_{m}\delta_{\eta}^{\rho}\delta_{\gamma}^{\kappa} + \zeta_{2} \mathbb{P}_{\mathbf{(3_{SL(2)})}}{}^{\rho}{}_{\eta}{}^{\kappa}{}_{\gamma} \delta_{n}^{p}\delta_{m}^{k} -\delta_{M}^{K}\delta_{N}^{P}\right) K_{k \kappa}
\end{align}
where $\zeta_{1}$ and $\zeta_{2}$ are two real parameters. The symmetric part of the generators of the gauge group, the intertwinning tensor, should be in the same representation whether or not we consider an $\mathbb{R}^{+}$ gauging. This is necessary in order to preserve the two-form field content of the theory \cite{diffon_supergravities_2009}. This is verified for $\zeta_{1} = -\zeta_{2} = 6$. The generators still have to verify a set of constraints which can be expressed in terms of the tensors introduced before as
\begin{align}
	\label{contraint1}
	0 =&X_{MN}{}^{P} K_{P} + 6\mathbb{P}_{\mathbf{(8)}}{}^{r}{}_{m}{}^{p}{}_{n}K_{r \gamma}K_{p \eta} - 6\mathbb{P}_{\mathbf{(3_{SL(2)})}}{}^{\delta}{}_{\gamma}{}^{\rho}{}_{\eta} K_{m \delta}K_{n \rho}-K_{m \gamma}K_{n \eta}\\
	\begin{split}
	\label{contraint2}	
	0 =& X_{PM}{}^{N}X_{NK}{}^{R}+X_{PK}^{N}X_{MN}{}^{R} - X_{PN}{}^{R}X_{MK}{}^{N}-K_{P}X_{MK}{}^{R}\\
	&+6\left(\mathbb{P}_{\mathbf{(8)}}{}^{q}{}_{p}{}^{n}{}_{m}\delta_{\rho}^{\sigma}\delta_{\gamma}^{\eta}-\mathbb{P}_{\mathbf{(3_{SL(2)})}}{}^{\sigma}{}_{\rho}{}^{\eta}{}_{\gamma}\delta_{p}^{q}\delta_{m}^{n}\right)K_{q \sigma}X_{n \eta, k \kappa}{}^{r \delta}\\
	&-6\left(\mathbb{P}_{\mathbf{(8)}}{}^{q}{}_{p}{}^{r}{}_{n}\delta_{\rho}^{\sigma}\delta_{\eta}^{\delta}-\mathbb{P}_{\mathbf{(3_{SL(2)})}}{}^{\sigma}{}_{\rho}{}^{\delta}{}_{\eta}\delta_{p}^{q}\delta_{n}^{r}\right)K_{q \sigma}X_{m \gamma, k \kappa}{}^{n \eta} \\
	&+6\left(\mathbb{P}_{\mathbf{(8)}}{}^{q}{}_{p}{}^{n}{}_{k}\delta_{\rho}^{\sigma}\delta_{\kappa}^{\eta}-\mathbb{P}_{\mathbf{(3_{SL(2)})}}{}^{\sigma}{}_{\rho}{}^{\eta}{}_{\kappa}\delta_{p}^{q}\delta_{k}^{n}\right)K_{q \sigma}X_{m \gamma, n \eta}{}^{r \delta}.
	\end{split}
\end{align}

\subsection{Generalised Dynamical Fluxes}

 Now that we have described the embedding tensor of maximal supergravity in 8 dimensions with a gauged trombone symmetry we look at the fluxes of $\mathbb{R}^{+}\times SL(3)\times SL(2)$ EFT. First let us consider the generalised metric of the extended space. We can define a generalised metric H living into the quotient $\mathbb{R}^{+}\times \frac{SL(3)}{SO(3)}\times \frac{SL(2)}{SO(2)}$ which transforms covariantly under $\mathbb{R}^{+}\times SL(3) \times SL(2)$ and is invariant under the maximal compact subgroup of $E_{3(3)}$ i.e $SO(3)\times SO(2)$. Due to the product structure of the group, we define a bein which splits as
\begin{equation}
\label{Bein}
	E_{\bar{A}}{}^{M} = e^{-\Delta} e_{\bar{a}}{}^{m} l_{\bar{\alpha}}{}^{\gamma}
\end{equation}
 where $\Delta$ is the $\mathbb{R}^{+}$ component of the metric, $e_{\bar{a}}{}^{m}$ and $l_{\bar{\alpha}}{}^{\gamma}$ the $SL(3)$ and $SL(2)$ beins respectively. $\bar{a}$ and $\bar{\alpha}$ are $SO(3)$ and $SO(2)$ planar indices respectively. The metric of the internal space is then
 \begin{equation}
 \label{metric}
 	H^{M N} = E_{\bar{A}}{}^{M} E_{\bar{B}}{}^{N}\delta^{\bar{A}\bar{B}}  = e^{-2 \Delta} H^{m n} g^{\gamma \eta}
 \end{equation}
 where
 \begin{align}
 \label{metrics}
 \begin{split}
 	H^{m n} &= e_{\bar{a}}{}^{m} e_{\bar{b}}{}^{n}\delta^{\bar{a}\bar{b}} \\
g^{\gamma \eta} &= l_{\bar{\alpha}}{}^{\gamma}l_{\bar{\beta}}{}^{\eta}\delta^{\bar{\alpha}\bar{\beta}} 
\end{split}
 \end{align}  correspond to an $SL(3)$ an $SL(2)$ metric respectively.\\

	Having defined the bein and a consistent generalised Lie derivative of the theory, one defines the generalised fluxes as
	\begin{equation}
		\mathcal{L}_{E_{\bar{A}}}E_{\bar{B}} = F_{\bar{A} \bar{B}}{}^{\bar{C}}E_{\bar{C}}.
	\end{equation}
 In a coordinate frame, we find the fluxes to be\footnote{For more details see \cite{aldazabal_extended_2013}.}

	\begin{equation}
	\label{fluxPro}
F_{MN}{}^{P} = \Omega_{MN}{}^{P}-(2\mathbb{P}_{\mathbf{(8,1)}}{}^{P}{}_{N}{}^{R}{}_{S} + 3\mathbb{P}_{\mathbf{(1,3)}}{}^{P}{}_{N}{}^{R}{}_{S}) \Omega_{RM}{}^{S} + \frac{1}{6} \Omega_{RM}{}^{R}\delta_{N}^{P}
\end{equation}
where
\begin{equation}
	\label{Weit}
	\Omega_{MN}{}^{P} = (E^{-1})_{N}{}^{\bar{A}} \partial_{M}E_{\bar{A}}{}^{P}
\end{equation} 
is the Weitzenböck connection\footnote{We are abusing notation as the Weitzenböck connection should be globally defined, which is a priori not the case here.}. Now, using the expressions of the bein \eqref{Bein} we obtain
\begin{align}
	\begin{split}
	\Omega_{M N}{}^{P} &= -\partial_{M}\Delta \delta_{N}^{P}+ \delta_{\eta}^{\rho} (e^{-1})_{n}{}^{\bar{a}}\partial_{m \gamma} (e_{\bar{a}}{}^{p})+ \delta_{n}^{p} (l^{-1})_{\eta}{}^{\bar{\alpha}} \partial_{m \gamma} (l_{\bar{\alpha}}{}^{\rho}) \\
	&= - \partial_{M}\Delta \delta_{N}^{P} + \delta_{\eta}^{\rho}\Omega_{m \gamma, n}{}^{p}+ \delta_{n}^{p} \Omega_{m \gamma, \eta}{}^{\rho}
	\end{split}
\end{align}
where 
\begin{align}
	\begin{split}
	\delta_{\eta}^{\rho}\Omega_{m \gamma, n}{}^{p}= \delta_{\eta}^{\rho} (e^{-1})_{n}{}^{\bar{a}}\partial_{m \gamma} (e_{\bar{a}}{}^{p}) \in \mathbf{(3,2)\times (8,1)} \\
	\delta_{n}^{p} \Omega_{m \gamma, \eta}{}^{\rho}= \delta_{n}^{p} (l^{-1})_{\eta}{}^{\bar{\alpha}} \partial_{m \gamma} (l_{\bar{\alpha}}{}^{\rho}) \in \mathbf{(3,2)\times (1,3)}
	\end{split}
\end{align} The first term $-\partial_{M}\Delta\delta_{N}^{P}$ obviously lives in $\mathbf{(3,2)\times (1,1) = (3,2)}$.\\

After some manipulations we find the following generalised fluxes
\begin{align}
	\begin{split}
	F_{MN}{}^{P} &= \left[f_{\gamma}{}^{p z}\epsilon_{z m n} -\frac{3}{4} \mathbb{P}_{\mathbf{(8)}}{}^{r}{}_{m}{}^{p}{}_{n} \xi_{r\gamma} \right] \delta_{\eta}^{\rho} +\left[\mathbb{P}_{\mathbf{(3_{SL(2)})}}{}^{\rho}{}_{\eta}{}^{\delta}{}_{\gamma}\xi_{m \delta}\right] \delta_{n}^{p}  \\
	&+\left(\frac{3}{2} -\frac{3}{4}\zeta\right)\mathbb{P}_{\mathbf{(8)}}{}^{r}{}_{m}{}^{p}{}_{n} K_{r \gamma} + \zeta \mathbb{P}_{\mathbf{(3_{SL(2)})}}{}^{\rho}{}_{\eta}{}^{\delta}{}_{\gamma}  K_{m \delta} - K_{m \gamma}\delta_{n}^{p} \delta_{\eta}^{\rho}
	\end{split}
\end{align}
where
\begin{align}
	\mathbf{(6,2)}&: f_{\gamma}{}^{p z} = \epsilon^{k q (z}\Omega_{k \gamma, q}{}^{p)}\\
	\mathbf{(3,2)}&:\left\lbrace \begin{array}{l}	
	\theta_{m \gamma} = \Omega_{r \gamma, m}{}^{r}-4\partial_{m \gamma}\Delta\\
	\tilde{\theta}_{m \gamma} = \Omega_{m \delta, \gamma}{}^{\delta}-3\partial_{m \gamma}\Delta\\
	K_{m \gamma} = -\frac{1}{6}(\theta_{m \gamma} + \tilde{\theta}_{m \gamma})\\
	\xi_{m \gamma} = (\tilde{\theta}_{m \gamma} - \theta_{m \gamma}) - \zeta K_{m \gamma} \end{array}\right.
\end{align}  
  and $\zeta$ is only used to write the fluxes in a similar form compared to the gauge generators \eqref{gaugeGenTrom}. Choosing $\zeta = -6$ gives us the the same expressions we found after considering the intertwinning tensor constraint in the context of D=8 gauged maximal supergravity with gauged trombone symmetry. We thus have to consider the quadratic constraints \eqref{contraint1} and \eqref{contraint2} on K and f. We present simplified expressions of these constraints for the type IIB supergravity solution of the section condition in the last section, after choosing an appropriate ansatz of the generalised bein \eqref{Bein}.
  
\section{Equations of motion}

	We will now look at the equations of motion of the theory. A general expression of these equations was obtained in \cite{coimbra_$e_dd$_2011,coimbra_supergravity_2012} using the supersymmetric variations of the internal and external gravitino and a torsion-free/metric compatible connection. In fact it is precised that in our case, for $E_{3(3)}$ EFT, one can define a unique generalised Christoffel symbol. In this section we find the expression of this generalised torsion free, metric compatible connection. We then find a generalised Ricci tensor following the construction of \cite{aldazabal_extended_2013} for $E_{7(7)}$ EFT. In the last section we finally obtain the equations of motion of type IIB supergravity after having chosen an appropriate ansatz for the generalised $\mathbb{R}^{+}\times SL(3)\times SL(2)$ bein.

	\subsection{Generalised Christoffel Symbol}

	Connections are defined to describe how a field is transported along curves on a manifold. Their definition can thus be chosen to be exactly the same as the one from Riemannian geometry
	\begin{equation}
		\label{connection}
		\nabla_{M}E_{\bar{A}}{}^{N} = \partial_{M}E_{\bar{A}}{}^{N} + \Gamma_{M K}{}^{N}E_{\bar{A}}{}^{K}.
	\end{equation}
	The torsion however is defined using the Lie derivative and will differ from the usual Riemannian geometry \cite{coimbra_$e_dd$_2011}
	
	\begin{equation}
		\mathcal{T}_{\bar{A}\bar{B}}{}^{\bar{C}} = (E^{-1})_{M}{}^{\bar{C}}\left(\mathcal{L}_{E_{\bar{A}}}^{\nabla} - \mathcal{L}_{E_{\bar{A}}}\right)E_{\bar{B}}{}^{M}
	\end{equation}
	with $\mathcal{L}^{\nabla}$ the generalised Lie derivative \eqref{diffZ} where every derivative is replaced by a covariant one. Requiring that the generalised torsion is null we get from \eqref{connection} that the generalised connection $\Gamma$ should verify the following generalised torsion condition	
\begin{equation}
	\label{torsionG}
	\Gamma_{MN}{}^{P} = 2 \mathbb{P}_{\mathbf{(8,1)}}{}^{P}{}_{N}{}^{D}{}_{Q} \Gamma_{DM}{}^{Q} + 3\mathbb{P}_{\mathbf{(1,3)}}{}^{P}{}_{N}{}^{D}{}_{Q} \Gamma_{DM}{}^{Q} - \frac{1}{6} \Gamma_{DM}{}^{D}\delta_{N}^{P}
\end{equation}
which can also be written
\begin{equation}
\label{torsionCond}
	2\Gamma_{[MN]}{}^{P} = -Z^{P}{}_{N}{}^{R}{}_{K}\Gamma_{R M}{}^{K}.
\end{equation}
Using those expressions it is possible to seek a generalised Christoffel symbol of the form\footnote{For more details see Appendix \ref{AnnexeB}.}

\begin{equation}
	\label{gamPresque}
		\Gamma_{MN}{}^{P} = \Gamma_{m \gamma, n}{}^{p} \delta_{\eta}^{\rho} + \Gamma_{m \gamma, \eta}{}^{\rho} \delta_{n}^{p} + \text{trace   terms}.
\end{equation}	
Now, considering the metric compatibility condition
\begin{equation}
	\label{metricCompatibility}
	0 = \nabla_{M}H^{NP} = \partial_{M} H^{NP} + \Gamma_{MR}{}^{N} H^{RP} + \Gamma_{MR}{}^{P} H^{RN}
\end{equation}
and the splitting of the metric \eqref{metric} the first two terms of the expression \eqref{gamPresque} are found to be
\begin{align}
	\label{gamma3d}
		\Gamma_{m \gamma, n}{}^{p} &= \frac{1}{2}H^{pr}\left( \partial_{m \gamma} H_{n r} + \partial_{n \gamma} H_{m r} - \partial_{r \gamma} H_{m n}\right)\\
	\label{gamma2d}
		\Gamma_{m \gamma, \eta}{}^{\rho} &= \frac{1}{2} H^{\rho \delta} \left( \partial_{m \gamma} H_{\eta \delta} + \partial_{m \eta} H_{\gamma \delta} - \partial_{m \delta} H_{\gamma \eta}\right).
\end{align}
The first term \eqref{gamma3d} is just 2 copies of a three dimensional usual Riemannian Christoffel symbol (for each value of $\gamma$), and the second term \eqref{gamma2d} is 3 copies of a two dimensional one (for each value of m).
Finally, using the torsion condition \eqref{torsionG} we find the generalised Christoffel symbol, with vanishing generalised torsion and metric compatibility to be
\begin{align}
	\begin{split}
	\label{chris}
	\Gamma_{M N}{}^{P} = \Gamma_{m \gamma n \eta}{}^{p \rho} &= \Gamma_{m \gamma, n}{}^{p} \delta_{\eta}^{\rho} + \Gamma_{m \gamma, \eta}{}^{\rho} \delta_{n}^{p} + 2\left(H^{p k}H_{m n}\partial_{k \gamma} \Delta \delta_{\eta}^{\rho} - \partial_{n \gamma} \Delta \delta_{m}^{p} \delta_{\eta}^{\rho}\right) \\
	&+3\left(H^{\rho \kappa}H_{\gamma \eta}\partial_{m \kappa} \Delta \delta_{n}^{p} - \partial_{m \eta}\Delta \delta_{n}^{p} \delta_{\gamma}^{\rho}\right) + \partial_{M} \Delta \delta_{N}^{P}
		\end{split}
\end{align}	
whose traces are
\begin{equation}
	\Gamma_{R M}{}^{R} = -\Gamma_{M R}{}^{R} = 6 \partial_{M} \Delta.
\end{equation}
This comes from the fact that the scalar that transforms properly under generalised diffeomorphisms is $e^{-6\Delta}$ for $SL(3)\times SL(2)$ i.e.
\begin{equation}
	\delta_{\xi}(e^{-6\Delta}) = \partial_{P}\left(e^{-6\Delta}\xi^{P}\right).
\end{equation}
We use the fact that the scalars of the theory should be of this particular form later in order to define a proper ansatz for the generalised metric and find the equations of motion one expects in F-theory.
	\subsection{Generalised Ricci Tensor}
A generalised Ricci tensor for the $\mathbb{R}^{+}\times E_{7(7)}$ EFT which transforms covariantly under generalised diffeormorphisms was proposed in \cite{aldazabal_extended_2013}. It seems to hold for any exceptional field theory as it is written in terms of the tensor Z without need of its precise form. Here we show that for $\mathbb{R}^{+}\times SL(3)\times SL(2)$ it gives the expected equations of motion, thus confirming the proposed form of a generalised Ricci tensor in exceptional field theory. We review the main steps in order to construct a generalised Ricci tensor.\\

The usual Riemann tensor of a Riemannian space can be expressed as
\begin{equation}
	\label{riemann}
	R_{MNP}{}^{R} = \partial_{M}\Gamma_{NP}{}^{R} - \partial_{N}\Gamma_{MP}{}^{R} + \Gamma_{ML}{}^{R}\Gamma_{NP}{}^{L} - \Gamma_{NL}{}^{R}\Gamma_{MP}{}^{L}.
\end{equation}
This object however does not transform properly under $SL(3)\times SL(2)$ generalised diffeomorphisms. Its non covariant variation is
\begin{align*}
	\Delta_{\xi}R_{MNP}{}^{R} =& 2\Delta_{\xi}\Gamma_{[MN]}{}^{Q}\Gamma_{QP}{}^{R}
\end{align*}
where $\Delta_{\xi} = \delta_{\xi} - \mathcal{L}_{\xi}$.
If one considers the torsion condition of the generalised Christoffel symbol \eqref{torsionCond}, the non covariant variation of the Riemann  tensor is null if Z = 0 i.e. if the usual torsion condition $\Gamma_{[MN]}{}^{P}$ = 0 is satisfied. Now, the usual Ricci tensor should be
	\begin{equation}
		R_{MN} = R_{M R N}{}^{R}
	\end{equation}
but again this does not transform as a tensor under generalised diffeomorphisms. Its  non covariant variation is
\begin{align}
\Delta_{\xi}R_{MP} =& 2\Delta_{\xi}\Gamma_{[MR]}{}^{Q}\Gamma_{QP}{}^{R}.
\end{align}
One can then construct the following generalised Ricci tensor
\begin{equation}
	\label{genRicci}
	\mathcal{R}_{MN} = \frac{1}{2}\left(R_{MN} + R_{NM} + \Gamma_{RM}{}^{P} Z^{RS}{}_{PQ} \Gamma_{SN}{}^{Q}\right)
\end{equation}
which verifies
\begin{equation}
	 \Delta_{\xi} \mathcal{R}_{M N} = 0.
\end{equation}
We will not detail here the expression of the generalised Ricci tensor obtained using our result on a generalised Christoffel symbol \eqref{chris}. This will be our focus in the last section, where we consider a particular ansatz for the $\mathbb{R}^{+}\times SL(3)\times SL(2)$ bein in terms of the fields of type IIB supergravity. 

\section{Recovering F-Theory}
	In this last section we use the results obtained before in order to relate $\mathbb{R}^{+}\times SL(3)\times SL(2)$ exceptional field theory to F-theory. We will begin by presenting a short review of F-theory, in the particular viewpoint of type IIB string theory with varying axio-dilaton. We then show that considering a non trivial solution to the section condition allows us to describe the monodromies of (p,q) 7-branes appearing in F-theory. Finally we consider an ansatz for the $\mathbb{R}^{+}\times SL(3)\times SL(2)$ bein which leads to the type IIB equations of motion. We also show that the equations of motion obtained on the internal space using the generalised Ricci tensor and the generalised Christoffel symbol are equivalent to the Ricci-flatness of a 4 dimensional usual Ricci tensor, of which two of the dimensions are fibered as one expects from F-Theory.
	
\subsection{Review of F-theory}

F-theory was constructed as a 12 dimensional geometrisation of type IIB string theory with 7-branes. In D=10 type IIB, the axion $C_{0}$ and the dilaton $\Phi$ can be arranged in a manifestly $SL(2,\mathbb{Z})$ complex field $\tau$ named the axio-dilaton. In F-theory this axio-dilaton is the complex parameter of a two dimensional torus, with constant volume, fibered over the 10 dimensional original space. This allows in particular to describe a varying axio-dilaton with respect to the directions normal to the D7 brane, thus taking into account the backreaction of the brane onto the geometry, and allowing for a strong coupling description of type IIB string theory \cite{weigand_lectures_2010, weigand_tasi_2018}. \\

Let us now go over some of the known results of F-theory. The effective bosonic action of type IIB supergravity theory in the Einstein frame is
\begin{equation}
\label{actionSII}
	\frac{1}{2\pi}S_{\text{IIB}} = \int \text{d}^{10}x\sqrt{-g}\left(R-\frac{\partial_{\mu}\tau\partial^{\mu}\bar{\tau}}{2\text{Im}(\tau)^{2}}-\frac{1}{2} \frac{|G_{3}|^{2}}{\text{Im}(\tau)}-\frac{1}{4}|F_{5}|^{2}\right)+\frac{1}{4i}\int \frac{1}{\text{Im}(\tau)}C_{4}+G_{3}\wedge \bar{G_{3}}
\end{equation}
where
\begin{align}
\begin{split}
\label{field}
	&\tau = C_{0} + ie^{-\phi}, \ \ \ \ G_{3} = dC_{2}-\tau dB_{2},\\
	 &F_{5} = dC_{4}-\frac{1}{2}C_{2}\wedge dB_{2}+\frac{1}{2} B_{2}\wedge dC_{2}, \ \ \ \ \ \ \ |F_{p}| = \frac{1}{p!}F_{\mu_{1}..\mu_{p}}F^{\mu_{1}..\mu_{p}}
	 \end{split}
\end{align} 
with $\tau$ the axio-dilaton, $B_{2}$ the NS-NS two form and $C_{p}$ the RR p forms. With this expression it is clear that the following $SL(2, \mathbb{R})$ transformations leave the action invariant
\begin{align}
	\label{monodromy}
	\tau \rightarrow \frac{a\tau + b}{c\tau+d}, \ \ \ \  \left(\begin{array}{c} C_{2} \\ B_{2} \end{array}\right) = M\left(\begin{array}{c} C_{2} \\ B_{2} \end{array}\right), \ \ \ \ C_{4} \rightarrow C_{4}, \ \ \ \ g_{\mu \nu} \rightarrow g_{\mu \nu}
\end{align}
where
 \begin{equation}
 	M = \left(\begin{array}{cc}  a & b \\ c & d \end{array}\right) \in SL(2, \mathbb{R}).
 \end{equation}
 One should note that the group $SL(2, \mathbb{R})$ is broken to $SL(2,\mathbb{Z})$  when one is considering non perturbative effects. Now, let us consider the simplest case first in flat space with a D7 brane along $\mathbb{R}^{1,7}\subset \mathbb{R}^{1,9}\simeq \mathbb{R}^{1,7}\otimes \mathbb{C}$. As the D7 brane is a magnetic charge for the axion $C_{0}$, one can show that, considering supersymmetry constraints, the axio-dilaton should behave as
 \begin{equation}
 	\tau(z) = \frac{1}{2\pi i}\text{ln}(z-z_{0})+ \text{terms regular at} \ z_{0}
 \end{equation}
 with $z$ the complex coordinate of the normal space to the brane and $z_{0}$ the position of the brane. This implies in particular that the dilaton transforms as
 \begin{equation}
 	\tau \rightarrow \tau+1
 \end{equation}
 as one encircles the brane around $z_{0}$, which corresponds to the $SL(2, \mathbb{Z})$ transformation introduced before with matrix parameter
 \begin{equation}
 	M = \left(\begin{array}{cc}  1 & 1 \\ 0 & 1 \end{array}\right).
 \end{equation}
 Now one can consider a more general 7-brane. A D7 brane corresponds to an object on which fundamental strings can end, strings which are coupled to the NS-NS two-form $B_{2}$. It is however possible to consider a BPS state of p fundamental strings, and q D1-strings called (p,q) strings, which couples to the field $pB_{2}+qC_{2}$. This leads to the consideration of another category of 7-branes, (p,q) 7-branes, on which (p,q) strings can end. It is then possible to show that the monodromies induced as one encircles a (p,q) 7-brane are given by
 
 \begin{equation}
 	\label{monodromyMatrix}
 	M = \left(\begin{array}{cc}  1+pq & p^{2} \\ -q^{2} & 1-pq \end{array}\right).
 \end{equation}

\subsection{Type IIB ansatz and generalised diffeomorphisms}
	In order to consider a type IIB solution of the $\mathbb{R}^{+}\times SL(3)\times SL(2)$ exceptional field theory the usual ansatz is \eqref{trivialAssump} i.e. that the fields only depend on the coordinates $X^{3 \gamma}$. This effectively leads to an 8+2 dimensional theory where the fields will have a dependency on the 8 dimensional external space time, and two coordinates of the six dimensional internal extended space.
Now, let us consider a particular choice of gauge for the generalised bein in terms of the fields of type IIB supergravity by breaking the $SL(3$) subgroup into  $SL(3)\rightarrow SL(2)\times U(1)$. The $SL(3)$ bein can be chosen to be
	\begin{equation}	
	\label{ansatz}
	e_{\hat{\bar{a}}}{}^{\hat{m}} =  \begin{pmatrix} e^{\frac{\Delta'}{2}} \begin{pmatrix} e^{\frac{\phi}{2}} & e^{\frac{\phi}{2}}C_{0} \\ 0 & e^{-\frac{\phi}{2}}\end{pmatrix}  &\begin{matrix} 0 \\ 0 \end{matrix} \\ \begin{matrix} e^{-\Delta'}B  & e^{-\Delta'} C \end{matrix} & e^{-\Delta'} \end{pmatrix}
	\end{equation}
where $\phi$ and $C_{0}$ are the dilaton and axion respectively. B, C are defined properly below in terms of $B_{2}$ and $C_{2}$ and $\Delta'$ will be related to the scale factor $\Delta$ introduced in equation \eqref{metric}. In order to understand what are the fields B and C we look at the action of a generalised Lie derivative \eqref{diffProj} of a generalised vector $V^{M} \equiv (V^{1 \gamma}, V^{2 \gamma}, v^{\gamma})$  onto the fields B and C when the usual solution to the section condition \eqref{trivialAssump} is verified\footnote{The expression of the action of generalised Lie derivatives onto the representation \textbf{(3,1)} can be found in \cite{hohm_tensor_2015}.}
\begin{align}
	\label{diffC}
 \mathcal{L}^{0}_{V}(C) &= v^{\gamma}\partial_{3 \gamma}C + \partial_{3 \gamma}v^{\gamma}C - \partial_{3 \gamma}V^{2\gamma}\\
 \label{diffB}
 \mathcal{L}^{0}_{V}(B) &= v^{\gamma}\partial_{3 \gamma}B + \partial_{3 \gamma}v^{\gamma}B - \partial_{3 \gamma}V^{1\gamma}.
\end{align}
 Moreover one can show that the antisymmetric 2 dimensional tensor $\epsilon^{\alpha \beta}$ is invariant under generalised diffeomorphisms. This allows to relate the \textbf{2} representation of $SL(2)$ to its dual $\mathbf{\bar{2}}$ using
 \begin{equation}
 	V_{\alpha} = V^{\beta}\epsilon_{\beta \alpha}, \ \ \ \ V^{\alpha} = \epsilon^{\alpha \beta} V_{\beta},
 \end{equation}
thus allowing us to rewrite the generalised Lie derivatives as
\begin{align}
 \mathcal{L}^{0}_{V}(C) &= v^{\gamma}\partial_{3 \gamma}C + \partial_{3 \gamma}v^{\gamma}C - \partial_{3 \gamma}V^{2}{}_{\eta}\epsilon^{\gamma \eta}\\
 \mathcal{L}^{0}_{V}(B) &= v^{\gamma}\partial_{3 \gamma}B + \partial_{3 \gamma}v^{\gamma}B - \partial_{3 \gamma}V^{1}{}_{\eta}\epsilon^{\gamma \eta}.
\end{align}
 To be more precise, we can define B and C to be the Hodge duals of the NS-NS ($B_{2}$) and RR ($C_{2}$) two-forms on the two dimensional space with metric $G_{\gamma \eta}\propto g_{\gamma \eta}$\footnote{In the last section we find that $G_{\gamma \eta} = e^{-6\Delta} g_{\gamma \eta}$ in order to recover the equations of motion of type IIB supergravity.}
\begin{align}
\begin{split}
		B &= \frac{1}{\sqrt{|G|}}\frac{\epsilon^{\gamma \eta}}{2}B_{\gamma \eta} \\
		C &= \frac{1}{\sqrt{|G|}}\frac{\epsilon^{\gamma \eta}}{2}C_{\gamma \eta}\\
		V^{1 \gamma} &= V^{1}{}_{\eta}\epsilon^{\gamma \eta} = \frac{\lambda^{B}{}_{\eta}}{\sqrt{|G|}}\epsilon^{\gamma \eta} \\
		V^{2 \gamma} &= V^{2}{}_{\eta}\epsilon^{\gamma \eta} = \frac{\lambda^{C}{}_{\eta}}{\sqrt{|G|}}\epsilon^{\gamma \eta}
\end{split}
\end{align}
therefore recovering the gauge transformations of $B_{2}$ and $C_{2}$ 
\begin{align}
	\begin{split}
	C_{\gamma \eta} \rightarrow C_{\gamma \eta} + \partial_{[\gamma}\lambda^{C}{}_{\eta]}\\
	B_{\gamma \eta} \rightarrow B_{\gamma \eta} + \partial_{[\gamma}\lambda^{B}{}_{\eta]}
	\end{split}
\end{align}
with $\lambda^{C}$ and $\lambda^{B}$ the one form parameters of the gauge transformations.\\

	As expected from an exceptional field theory we can see that using generalised diffeomorphisms, we can describe the usual diffeomorphisms of the two dimensional space (via $v^{\gamma}$), combined with two gauge transformations of the RR and NS-NS two-forms (via $V^{1 \gamma}$ and $V^{2 \gamma}$). In the next section we look at the implication of a more general solution to the constraint, which allows one to describe the monodromies of (p,q) 7-branes using the generalised Lie derivatives.
\subsection{F-theory as $\mathbb{R}^{+}\times E_{3(3)}$ EFT with non standard solution to the section condition}

One can solve the section condition \eqref{strongConstraint} by requiring that the fields depend on two coordinates, but allow them to be a combination of the ordinary ones $X^{3 \gamma}$ and the ones associated to winding and D1 brane wrapping $X^{\hat{m} \gamma}$.
We propose the following solutions to the section condition of $\mathbb{R}^{+}\times SL(3)\times SL(2)$ exceptional field theory
\begin{equation}
	\label{assumption}
	A = f(X^{3 \gamma} + A_{\hat{m}}X^{\hat{m} \gamma})
\end{equation}
for any field A and with $\hat{m}$=1,2. The terms $A_{\hat{m}}$ are constant with respect to the 6 dimensional internal space, but can a priori depend on the 8 dimensional space time coordinates. We recover the usual section condition for $A_{\hat{m}} = 0$. One should note that the structure of the ansatz \eqref{assumption} as a global function of the combined coordinates is necessary in order for products and inverts of fields to be well defined i.e. so that they are also solutions to the section condition. \\

Let us see what happens when performing generalised diffeomorphisms. As we are only interested in the extra terms compared to the usual section condition solution \eqref{trivialAssump}, we will denote by $\mathcal{L}^{0}_{\Lambda}$ the generalised Lie derivative with respect to a generalised vector V when $\partial_{1 \alpha}(A) = \partial_{2 \alpha}(A) = 0$ for any field A. This corresponds to the equations \eqref{diffC} and \eqref{diffB}.
Let us first look at the generalised Lie derivative of the fields $\phi$ and $C_{0}$
\begin{align}
	\begin{split}
	\mathcal{L}_{V}(e^{\phi}) &= \mathcal{L}^{0}_{V}(e^{\phi})+ V^{\hat{k} \kappa}\partial_{\hat{k} \kappa}(e^{\phi})\\& -e^{\phi}\partial_{1 \kappa}V^{1 \kappa}-e^{\phi}C_{0}\partial_{2 \kappa}V^{1 \kappa}+\left(\lambda(e^{\phi})+\frac{2}{3}\right)\partial_{\hat{k} \kappa}V^{\hat{k} \kappa}e^{\phi}\\
	\mathcal{L}_{V}(C_{0}) &= \mathcal{L}^{0}_{V}(C_{0}) + V^{\hat{k} \kappa}\partial_{\hat{k} \kappa}(C_{0}) -\partial_{1 \kappa}V^{2 \kappa}\\&- C_{0}\left(\partial_{2 \kappa}V^{2 \kappa}-\partial_{1 \kappa}V^{1 \kappa}\right)-C_{0}^{2}\partial_{2 \kappa}V^{1 \kappa}+ \lambda(C_{0})\partial_{\hat{k} \kappa}V^{\hat{k} \kappa}C_{0}.
	\end{split}
\end{align}
We consider generalised diffeomorphisms that satisfy $\partial_{\hat{k} \kappa}V^{\hat{k} \kappa} = V^{\hat{k} \kappa} \partial_{\hat{k} \kappa} = 0$ which can be achieved by considering that
\begin{equation}
\label{condition1}
A_{2}V^{2 \kappa} = -A_{1}V^{1 \kappa}.
\end{equation}
This gives
\begin{align}
	\begin{split}
	\mathcal{L}_{V}(e^{\phi}) &= \mathcal{L}^{0}_{V}(e^{\phi}) -e^{\phi}\partial_{1 \kappa}V^{1 \kappa}-e^{\phi}C_{0}\partial_{2 \kappa}V^{1 \kappa}\\
	\mathcal{L}_{V}(C_{0}) &= \mathcal{L}^{0}_{V}(C_{0}) - \partial_{1 \kappa}V^{2 \kappa}-C_{0}\left(\partial_{2 \kappa}V^{2 \kappa}-\partial_{1 \kappa}V^{1 \kappa}\right)-C_{0}^{2}\partial_{2 \kappa}V^{1 \kappa}.
	\end{split}
\end{align}
Here we see that the term $-\partial_{1 \kappa}V^{2 \kappa}$ is producing a shift of the axion, as is expected from a monodromy of a (p,0) 7-brane given by the equation \eqref{monodromy} and using the corresponding monodromy matrix \eqref{monodromyMatrix}. As the other terms are not particularly clear when one looks at the fields $\phi$ and $C_{0}$ let us consider the generalised Lie derivatives of the fields B and C. Considering the ansatz \eqref{condition1} we obtain
\begin{align}
	\mathcal{L}_{V}(C) &= \mathcal{L}^{0}_{V}(C) - B\partial_{1 \kappa}V^{2 \kappa} - C \partial_{2 \kappa}V^{2 \kappa}\\
	\mathcal{L}_{V}(B) &= \mathcal{L}^{0}_{V}(B) - B\partial_{1 \kappa}V^{1 \kappa} - C\partial_{2 \kappa}V^{1 \kappa}.
\end{align}
Now, to make sense of the two previous equation in terms of monodromies we take each component of $V^{M}$ to be linear in its coordinates. Requiring the conditions \eqref{assumption} and \eqref{condition1} we are able to recover the monodromies of a general (p,q) 7-brane encoded into the generalised Lie derivatives of the exceptional field theory
\begin{align}
	\begin{split}
	\mathcal{L}_{V}(C) &= \mathcal{L}^{0}_{V}(C) +pq C + p^{2} B \\
	\mathcal{L}_{V}(B) &= \mathcal{L}^{0}_{V}(B) -pq B - q^{2}C
	\end{split}
\end{align}
with the additional conditions
\begin{align}
	\begin{split}
	\partial_{1 \kappa}V^{1 \kappa} &= -\partial_{2 \kappa}V^{2 \kappa} = pq\\
	\partial_{1 \kappa}V^{2 \kappa} &= -p^{2}\\
	\partial_{2 \kappa}V^{1 \kappa} &= q^{2}\\
	qA_{1} &= pA_{2}.
	\end{split}
\end{align}
Now let us look at the particular case of a stack of p D7 branes, as an arbitrary (p',q') 7-brane can be mapped locally to a (p,0) one, using an $SL(2, \mathbb{Z})$ transformation. We can make the following ansatz for the dependency of the Lie derivative generalised vector parameter 
\begin{equation}
	V^{M} = (0, X^{3 \gamma} - \frac{p^{2}}{2}X^{1 \gamma}, 0)
\end{equation}
where we put $V^{3 \gamma}$ to zero to remove the diffeomorphisms component.
Using this we obtain the full transformations of the fields to be
\begin{align}
	\begin{split}
	\mathcal{L}_{V}(e^{\phi}) &= 0\\
	\mathcal{L}_{V}(C_{0}) &= p^{2}\\
	\mathcal{L}_{V}(C) &= p^{2} B + 2\\
	\mathcal{L}_{V}(B) &= 0
	\end{split}
\end{align}
The additional shift term in the action of the monodromy on C is coming from a breaking of the gauge symmetry invariance of this field, which is also the case for B. The gauge invariances of the fields B and C seem to be entirely constrained by the monodromies as one goes around a D7 brane. One could notice that the breaking of the gauge invariances is expected when non perturbative effects of string theory are taking into account. This is however not an acceptable explanation in our case as we are in the perturbative regime. A more appropriate explanation would be that the monodromies we described before have an interpretation only when one is considering that the only term appearing in the Chern-Simons action of a D7 brane is the $C_{8}$ term dual to the axion $C_{0}$ \cite{bergshoeff_seven-branes_2007}. The full Chern-Simons action is however
\begin{equation}
	\int_{\mathcal{M}_{8}} \mathbf{C}\wedge e^{-B_{2}}
\end{equation}
 where $\mathcal{M}_{8}$ is the brane world volume and 
 \begin{equation}
 	\mathbf{C} = \sum_{p=0..4} C_{2p}.
 \end{equation}
This might break the gauge invariances of both $B_{2}$ and $C_{2}$.
	\subsection{Equations of motion via the Generalised Ricci Tensor}			
	
	In this section we conclude by writing explicitly the equations of motion using the generalised Ricci tensor \eqref{genRicci} and with the help of the symbolic computer algebra system Cadabra \cite{peeters_eld-theory_nodate,peeters_symbolic_nodate}. To begin with let us consider the proposed ansatz \eqref{ansatz} for the generalised bein. As we showed that the scalars which transform properly are of the form $e^{6n\Delta}$ where $n \in\mathbb{Z}$, and as the $\mathbb{R}^{+}$ factor in the metric are of the form $e^{-2\Delta + \Delta'}$ and $e^{-2\Delta -2\Delta'}$, a plausible ansatz on the scalar $\Delta'$ is
	\begin{equation}
	\label{delta}
	\Delta'=-4\Delta.
	\end{equation}
Now according to \cite{coimbra_$e_dd$_2011}, the equations of motions should live in the representation
\begin{equation}
	\mathbf{(5,1)+(1,2)+(1,1)}.
\end{equation}
This leads to consider the following equations of motion
\begin{align}
	\begin{split}
	\label{eom}
	0 = \mathcal{R}_{m \gamma, n \eta}H^{m n} &\in \mathbf{(1,2)}\\
	0 = \mathcal{R}_{(m| \gamma, |n) \eta}g^{\gamma \eta} &\in \mathbf{(5,1)+(1,1)}.
	\end{split}
\end{align}
Using the definition of the generalised Ricci tensor \eqref{genRicci}, the generalised Christoffel symbol \eqref{chris}, the ansatz on the bein \eqref{ansatz} as well as the ansatz on $\Delta'$ \eqref{delta} we obtain the equations of motion of type IIB supergravity in 2 dimensions\footnote{The expression \eqref{eomGraviton} can differ from the litterature by a minus sign due to the definition of the Riemann tensor \eqref{riemann}.}
\begin{align}
	\label{eomGraviton}
	R_{\gamma \rho}\left[e^{-6\Delta}g_{\cdot \cdot}\right] + \frac{1}{2}\partial_{\gamma}\phi\partial_{\rho}\phi + \frac{1}{2}\partial_{\gamma}C_{0}\partial_{\rho}C_{0} &= 0\\
	\label{eomDilaton}
	g^{\gamma \rho}\left(\nabla_{\gamma}\nabla_{\rho}\phi-e^{2\phi}\nabla_{\gamma}C_{0}\nabla_{\rho}C_{0}\right) &= 0\\
	\label{eomAxion}
	g^{\gamma \rho}\left(\nabla_{\gamma}C_{0}\nabla_{\rho}C_{0} + 2\nabla_{\gamma}C_{0}\nabla_{\rho}\phi\right) &= 0.
\end{align}
where  $\partial_{\gamma} \equiv \partial_{3 \gamma}$ and $R_{\gamma \rho}[e^{-6\Delta}g_{\cdot \cdot}]$ corresponds to the usual Ricci tensor associated to the metric $e^{-6\Delta}g_{\gamma \rho}$. $\nabla$ is the covariant derivative whose connection is the usual two dimensional Christoffel symbol of the same metric. 
One should note that the only way to recover the equations of motions of type IIB supergravity is to combine the warp factor $\Delta$ with the $\mathbb{R}^{+}$ factor $\Delta'$ coming from the breaking of $SL(3)$ into $SL(2)\times U(1)$ as in \eqref{delta}. Finally as we stated before the fields should verify different constraints due to the quadratic conditions \eqref{contraint1} and \eqref{contraint2} which in the end can be recast into
\begin{equation}
\label{condFlux}
\tilde{\theta}_{3 [\gamma|}\Omega_{3 |\rho], k}{}^{r} = 0.
\end{equation}
If we write the geometric fluxes of the two dimensional space with bein $\tilde{l}_{\bar{\alpha}}{}^{\gamma} = e^{3\Delta}l_{\bar{\alpha}}{}^{\gamma}$ as
\begin{equation}
	w_{\gamma \eta}{}^{\rho} = 2(\tilde{l}^{-1})_{[\gamma|}{}^{\bar{\alpha}}\partial_{|\eta]}\tilde{l}_{\bar{\alpha}}{}^{\rho}
\end{equation}
we find that the condition \eqref{condFlux} is equivalent to
\begin{align}
	\begin{split}
	w_{[\gamma| \delta}{}^{\delta}\partial_{|\eta]}\Phi = 0&\\
	w_{[\gamma| \delta}{}^{\delta}\partial_{|\eta]}C_{0} = 0&\\
	w_{[\gamma| \delta}{}^{\delta}\partial_{|\eta]}\Delta = 0.
\end{split}
\end{align}
These can be solved in particular if we consider the trace $w_{\gamma \delta}{}^{\delta}$ to be null, which is equivalent to $w = 0$ for a two dimensional space. This in particular ensures that the two dimensional internal space is compact \cite{andriot_beta-supergravity:_2013}.\\

To conclude we show that the equations of motion \eqref{eom} are equivalent to the Ricci-flatness of a 4 dimensional space: a two torus with constant volume equal to one, fibered over a two dimensional Riemann space. To do that we consider a 4 dimensional space whose metric is
\begin{align}
	H_{M N} =& \begin{pmatrix}
H_{\hat{m} \hat{n}} & 0 \\ 
 0 & g_{\gamma \rho}
	\end{pmatrix}
\end{align}
where $M = (\hat{m}, \gamma)$  and with $\hat{m}$ and $\gamma$ being 1 or 2. $H_{\hat{m} \hat{n}}$ is an SL(2) metric while $g_{\gamma \rho}$ is a GL(2) one. Now in order to describe a fibration we will consider that every field only depends on the two coordinates $x^{\gamma}$. Considering the usual Riemannian Ricci tensor of this four dimensional space with $\partial_{\hat{m}} = 0$ we have for $M=\gamma$
\begin{align}
	R_{M=\gamma, P= \rho} = R_{\gamma \rho}-\frac{1}{4}\partial_{\gamma}{H^{k r}}\partial_{\rho}{H_{k r}}
\end{align}
where $R_{\gamma \rho}$ is the two dimensional usual Ricci tensor associated to the metric $g_{\gamma \rho}$. Assuming that the $SL(2)$ metric is of the usual form
\begin{equation}
	H_{\hat{m} \hat{n}} = \frac{1}{\text{Im}(\tau)} \left( \begin{array}{cc} |\tau|^{2} & -\text{Re}(\tau) \\ -\text{Re}(\tau) & 1 \end{array} \right)
\end{equation}
with the axio-dilaton $\tau$ given by \eqref{field}, we recover the expected equations of motion we derived before. The equations of motion for the dilaton \eqref{eomDilaton} and the axion \eqref{eomAxion} are obtained by considering the other components of the Ricci tensor $R_{M = m, P =p}$.

\section{Conclusion and outlook}
In this paper we showed that the equations of motion of F-theory as a Ricci-flatness of a four dimensional space with two of its dimensions fibered are obtained from the $\mathbb{R}^{+} \times SL(3)\times SL(2)$ exceptional field theory with the standard solution to the section condition. This was done by considering an appropriate ansatz \eqref{ansatz} on the generalised bein, a manifestly $\mathbb{R}^{+}\times SL(3)\times SL(2)$ covariant generalised Ricci tensor \eqref{genRicci} proposed in \cite{aldazabal_extended_2013} and a generalised Christoffel symbol \eqref{chris} with vanishing generalised torsion and metric compatibility. We also showed that one can solve the section condition of exceptional field theory and still allow the fields to have a dependency on the additional stringy coordinates. 
Every product and invert of fields satisfying our ansatz \eqref{assumption} still verifies the section condition and the ansatz therefore seems to be appropriate. Finally we showed that this new solution to the section condition describes the monodromies of (p,q) 7-branes in the context of F-theory. This leads to the breaking of the gauge invariances of the NS-NS and RR two-forms which could be explained by the fact that the monodromies \eqref{monodromy} are only valid when one is considering that the only field living on the world volume of the D7 brane is the RR $C_{8}$ form, not including other RR p-forms or NS-NS two-form. Interesting prospects would be to understand this gauge symmetry breaking, and to look at the equations of motion obtained from the generalised Ricci tensor but this time with a non standard solution to the section condition. 
\section*{Acknowledgements}
	I would like to express my gratitude to my PhD supervisor Mariana Graña for her guidance all along this project. I would also like to thank Iosif Bena, Pierre Heidmann, Ruben Monten, Edvard T. Musaev, Henning Samtleben, Raffaele Savelli, Charles Strickland-Constable and Daniel Waldram for helpfull discussions. This work was supported in part by the Ecole Doctorale Physique en Île-de-France fellowship and the ERC Consolidator Grant 772408-Stringlandscape.
\appendix

	\section{Projectors}
	\label{Projector}
	Here we present the construction of the projectors on the useful representations we used along this paper. A detailed construction can be found in \cite{de_wit_lagrangians_2003}: it is shown in the first appendix of this paper that for an arbitrary simple group G, with the exception of $E_{8}$, one can decompose the product of the fundamental representation of the group G ($\mathbf{D(\Lambda)}$) with its adjoint $\mathbf{Adj(G)}$ as
	\begin{equation}
	\label{spaces}
		\mathbf{D(\Lambda)\times Adj(G)\rightarrow D(\Lambda) + D_{1} + D_{2}}
	\end{equation}
where $\mathbf{D_{1}}$ and $\mathbf{D_{2}}$ are two other representations. The ones of interest for us are only the fundamental and the representation with the smaller dimension ($\mathbf{D_{1}}$), as they are the only representations allowed for the embedding tensor after one considers the linear constraint coming from supersymmetry consideration. If we note M the fundamental representation of G and $\lbrace t^{\alpha} \rbrace$ ($\alpha$ = 1...dim(G)) the generators of the adjoint of G, the projectors on those two representations can be written\footnote{The adjoint indices are raised and lowered using the Cartan-Killing metric.}
\begin{align}
	\begin{split}
	\mathbb{P}_{(D(\Lambda)) M}{}^{\alpha, N}{}_{\beta} &= A(t^{\alpha}t_{\beta})_{M}{}^{N} \\
	\mathbb{P}_{(D_{1}) M}{}^{\alpha, N}{}_{\beta} &= a_{1}\delta^{\alpha}{}_{\beta}\delta_{M}{}^{N} + a_{2}(t_{\beta}t^{\alpha})_{M}{}^{N} + a_{3}(t^{\alpha}t_{\beta})_{M}{}^{N}
	\end{split}
\end{align}
with A, $a_{i}$ constants which are given in \cite{de_wit_lagrangians_2003} for every simple groups.

Now let us look at the two simple groups of interest to us, $SL(3)$ and $SL(2)$, whose fundamental representations $\mathbf{(3)}$ and $\mathbf{(2)}$ are written m and $\gamma$ respectively. For clarity we will note $\lbrace t^{\alpha} \rbrace$ ($\alpha$ = 1,...,8) the generators of the adjoint of $SL(3)$, and $\lbrace s^{\tilde{\alpha}} \rbrace$ ($\tilde{\alpha}$ = 1,2,3) the ones of $SL(2)$. With this we find the projectors onto the the fundamental representation of $SL(3)$ and $D_{1} = \mathbf{(6)}$ to be
\begin{align}
\begin{split}
	\mathbb{P}_{\mathbf{(3)}, m}{}^{\alpha, n}{}_{\beta} &= \frac{3}{8}(t^{\alpha}t_{\beta})_{m}{}^{n}\\
	\mathbb{P}_{\mathbf{(6)}, m}{}^{\alpha, n}{}_{\beta} &= \frac{1}{2}\delta^{\alpha}{}_{\beta}\delta_{m}{}^{n} - \frac{1}{2}(t_{\beta}t^{\alpha})_{m}{}^{n} - \frac{1}{4} (t^{\alpha}t_{\beta})_{m}{}^{n}.
	\end{split}
\end{align}
For the $SL(2)$ case, the result is a little peculiar as one has the following relation
\begin{equation}
	\delta_{\tilde{\beta}}^{\tilde{\alpha}}\delta_{\gamma}^{\eta} - (s_{\tilde{\beta}}s^{\tilde{\alpha}})_{\gamma}{}^{\eta} - (s^{\tilde{\alpha}}s_{\tilde{\beta}})_{\gamma}{}^{\eta} = 0.
\end{equation}
The only representations left are then $\mathbf{D(\Lambda)} = \mathbf{(2)}$ and $\mathbf{D_{2}} = \mathbf{(4)}$. The projection onto the fundamental is
\begin{align}
	\begin{split}
	\mathbb{P}_{\mathbf{(2)}, \gamma}{}^{\tilde{\alpha}, \eta}{}_{\tilde{\beta}} &= \frac{2}{3}(s^{\tilde{\alpha}}s_{\tilde{\beta}})_{\gamma}{}^{\eta}\\
	&=\frac{2}{3}(\delta^{\tilde{\alpha}}{}_{\tilde{\beta}}\delta_{\gamma}^{\eta}-(s_{\tilde{\beta}}s^{\tilde{\alpha}})_{\gamma}{}^{\eta}).
	\end{split}
\end{align}
We write these projectors in the fundamental representation of each groups, leading for $SL(3)$ to
\begin{align}
	\begin{split}
	\mathbb{P}_{\mathbf{(3)} m n}{}^{p, a b}{}_{c} &= \frac{3}{8}(\delta_{m}^{p}\delta_{c}^{a}\delta_{n}^{b} - \frac{1}{3}\delta_{n}^{p}\delta_{c}^{a} \delta_{m}^{b} - \frac{1}{3}\delta_{m}^{p} \delta_{n}^{a}\delta_{c}^{b} +\frac{1}{9} \delta_{n}^{p} \delta_{m}^{a} \delta_{c}^{b})\\
	\mathbb{P}_{\mathbf{(6)}}{}_{mn}{}^{p,ab}{}_{c} &= \frac{1}{2} \epsilon_{mnr}\epsilon^{ab(r}\delta_{c}^{p)}
	\end{split}
\end{align}
and for $SL(2)$
\begin{align}
	\mathbb{P}_{\mathbf{(2)} \gamma \eta}{}^{\rho, \alpha \beta}{}_{\xi} = \frac{2}{3}(\delta_{\gamma}^{\rho}\delta_{\xi}^{\alpha}\delta_{\eta}^{\beta} - \frac{1}{2}\delta_{\eta}^{\rho}\delta_{\xi}^{\alpha} \delta_{\gamma}^{\beta} - \frac{1}{2}\delta_{\gamma}^{\rho} \delta_{\eta}^{\alpha}\delta_{\xi}^{\beta} +\frac{1}{4} \delta_{\eta}^{\rho} \delta_{\gamma}^{\alpha} \delta_{\xi}^{\beta}).
\end{align}
These expressions are found using the projectors on the adjoint $\mathbf{(8)}$ of $SL(3)$
\begin{equation}
	\mathbb{P}_{\mathbf{(8)}}{}^{m}{}_{n}{}^{p}{}_{q} = (t_{\alpha})_{n}{}^{m}(t^{\alpha})_{q}{}^{p} = \delta_{q}^{m}\delta_{n}^{p}-\frac{1}{3}\delta_{n}^{m}\delta_{q}^{p}
\end{equation}
and the adjoint $\mathbf{(3_{SL(2)})}$ of $SL(2)$
\begin{equation}
	\mathbb{P}_{\mathbf{(3_{SL(2)})}}{}^{\gamma}{}_{\eta}{}^{\rho}{}_{\delta} = (s_{\tilde{\alpha}})_{\eta}{}^{\gamma}(s^{\tilde{\alpha}})_{\delta}{}^{\rho} =  \delta_{\delta}^{\gamma}\delta_{\eta}^{\rho}-\frac{1}{2}\delta_{\eta}^{\gamma}\delta_{\delta}^{\rho}.
\end{equation}
\section{Determination of $\Gamma$}
	\label{AnnexeB}
	The expression of the generalised Christoffel symbol \eqref{chris} was hinted by a series of projections applied to the torsion condition \eqref{torsionG}. Here we detail the different relations that permitted in the end to look for a generalised Christoffel of the form \eqref{gamPresque}.
	
	First of all, one can relate the traces of the Christoffel symbol by taking the trace of the torsion condition
	\begin{equation}
	\Gamma_{MD}{}^{D} = -\Gamma_{DM}{}^{D}.
\end{equation}
By taking the partial traces on the different subspaces it is also possible to write the following relations
\begin{align}
	\begin{split}
	\Gamma_{m \gamma, n \eta}{}^{n \rho} &= 3 \Gamma_{n \eta, m \gamma}{}^{n \rho} -2\Gamma_{R M}{}^{R} \delta_{\eta}^{\rho} \\
	\Gamma_{m \gamma, n \eta}{}^{p \eta} &= 2 \Gamma_{n \eta, m \gamma}{}^{p \eta} - \Gamma_{R M}{}^{R} \delta_{n}^{p}
	\end{split}
\end{align}
which can be recast into
\begin{align}
	\begin{split}
	\Gamma_{r \gamma, m \delta}{}^{r \delta} &= \Gamma_{r\delta, m \gamma}{}^{r \delta}\\
	\Gamma_{m \delta, r \gamma}{}^{r \delta} &= \Gamma_{r \delta, m \gamma}{}^{r\delta}.
	\end{split}
\end{align}
Other useful relations are obtained by taking the projection of the torsion condition onto the representations $\mathbf{(8,1)}$ and $\mathbf{(1,3)}$
\begin{align}
	\begin{split}
	\mathbb{P}_{(8,1)}{}^{R}{}_{S}{}^{B}{}_{C} \Gamma_{AB}{}^{C} &= 2\mathbb{P}_{(8,1)}{}^{R}{}_{S}{}^{B}{}_{C} \Gamma_{BA}{}^{C} \\
	\mathbb{P}_{(1,3)}{}^{R}{}_{S}{}^{B}{}_{C} \Gamma_{AB}{}^{C} &= 3\mathbb{P}_{(1,3)}{}^{R}{}_{S}{}^{B}{}_{C} \Gamma_{BA}{}^{C}.
	\end{split}
\end{align} 
We also have to recall from \eqref{spaces} that
\begin{equation}
	\mathbb{P}_{\mathbf{D(\Lambda)}} + \mathbb{P}_{\mathbf{D_{1}}} +\mathbb{P}_{\mathbf{D_{2}}} = \text{Id}_{\mathbf{D(\Lambda)}\times \mathbf{Adj(G)}}
\end{equation}
which for the groups SL(2) and SL(3) can be written\footnote{In the case of $SL(2)$ the space $D_{1}$ is empty.}
\begin{align}
	\begin{split}
		\mathbb{P}_{\mathbf{(2)}} + \mathbb{P}_{\mathbf{(4)}} = \text{Id}_{\mathbf{(2)}\times \mathbf{(3)}}\\
		\mathbb{P}_{\mathbf{(3)}} + \mathbb{P}_{\mathbf{(6)}} +\mathbb{P}_{\mathbf{(15)}} = \text{Id}_{\mathbf{(3)}\times \mathbf{(8)}}
	\end{split}
\end{align}
where the traces on the spaces $\mathbf{(4)}$ and $\mathbf{(15)}$ are null.
The torsion condition can than be recast as
\begin{align}
	\begin{split}
	\left(\tilde{\Gamma}_{(15,2)} + \tilde{\Gamma}_{(3,4)}\right)_{MN}{}^{P} &= \Gamma_{MN}{}^{P} + \frac{1}{6} \Gamma_{DM}{}^{D}\delta_{N}^{P} - \frac{21}{24} \mathbb{P}_{(8,1)}{}^{K}{}_{M}{}^{P}{}_{N} \Gamma_{RK}{}^{R} - \frac{14}{6} \mathbb{P}_{(1,3)}{}^{K}{}_{M}{}^{P}{}_{N} \Gamma_{RK}{}^{R} \\
	&+\left[\frac{9}{4} \mathbb{P}_{(8,1)}{}^{K}{}_{M}{}^{P}{}_{N} \mathbb{P}_{(1,3)}{}^{R}{}_{K}{}^{S}{}_{T} + 4\mathbb{P}_{(1,3)}{}^{K}_{M}{}^{P}{}_{N} \mathbb{P}_{(8,1)}{}^{R}{}_{K}{}^{S}{}_{T} \right] \Gamma_{SR}{}^{T}.
	\end{split}
\end{align}
The relations
\begin{align}
	\begin{split}
	\frac{9}{4} \mathbb{P}_{(8,1)}{}^{K}{}_{M}{}^{P}{}_{N} \mathbb{P}_{(1,3)}{}^{R}{}_{K}{}^{S}{}_{T} \Gamma_{SR}{}^{T} &= \frac{3}{8} \mathbb{P}_{(8,1)}{}^{K}{}_{M}{}^{P}{}_{N} \Gamma_{SK}{}^{S} \\
	4\mathbb{P}_{(1,3)}{}^{K}{}_{M}{}^{P}{}_{N} \mathbb{P}_{(8,1)}{}^{R}{}_{K}{}^{S}{}_{T} \Gamma_{SR}{}^{T} &= \frac{4}{3} \mathbb{P}_{(1,3)}{}^{K}{}_{M}{}^{P}{}_{N} \Gamma_{SK}{}^{S}
	\end{split}
\end{align}
permit to write the last term between brackets into a trace part, which leads to
\begin{equation}
\Gamma_{MN}{}^{P} = \left(\tilde{\Gamma}_{(15,2)} + \tilde{\Gamma}_{(3,4)}\right)_{MN}{}^{P}  + \text{trace \ terms}.
\end{equation}
Using
\begin{align}
	\begin{split}
	\tilde{\Gamma}_{(15,2) m \gamma, n \eta}{}^{p \rho} &= \left[\mathbb{P}_{(15,2)}\Gamma\right]_{MN}{}^{P} = \left[\mathbb{P}_{(15)}\Gamma\right]_{m \gamma, n}{}^{p} \delta_{\eta}^{\rho} = \tilde{\Gamma}_{(15) m\gamma, n}{}^{p} \delta_{\eta}^{\rho} \\
	\tilde{\Gamma}_{(3,4) m \gamma, n \eta}{}^{p \rho} &= \left[\mathbb{P}_{(3,4)} \Gamma\right]_{MN}{}^{P} = \left[\mathbb{P}_{(15)}\Gamma\right]_{m \gamma, \eta}{}^{\rho} \delta_{n}^{p} = \tilde{\Gamma}_{(4) m \gamma, \eta}{}^{\rho} \delta_{n}^{p}
	\end{split}
\end{align}
with partial traces of $\tilde{\Gamma}_{\mathbf{(15)}}$ and $\tilde{\Gamma}_{\mathbf{(4)}}$ null, we have
\begin{equation}
\Gamma_{MN}{}^{P} = \tilde{\Gamma}_{\mathbf{(15)} m \gamma, n}{}^{p}\delta_{\eta}^{\rho} + \tilde{\Gamma}_{\mathbf{(4)} m \gamma, \eta}{}^{\rho}\delta_{n}^{p} + \text{trace \ terms}.
\end{equation}
Using this expression in the metric compatibility condition \eqref{metricCompatibility} and the torsion condition \eqref{torsionCond} leads to the solution \eqref{chris}.

\bibliography{Bibli}
\bibliographystyle{unsrt}
\end{document}